\documentclass[12pt]{article}
\pdfoutput=1
\usepackage{hyperref}
\usepackage{graphicx}
\usepackage{subcaption}
\graphicspath{{./picture/}}
\usepackage{empheq}
\usepackage{amssymb}
\usepackage{mathtools}
\usepackage{commath}
\usepackage{verbatim}
\usepackage{xcolor}
\usepackage{soul}
\usepackage{booktabs}
\usepackage{amsmath}
\usepackage{hhline}
\usepackage[scale={0.75,0.8}]{geometry}
\usepackage{url}
\usepackage{caption}
\usepackage[numbers, sort&compress]{natbib}
\usepackage{epsfig}
\usepackage{lscape}
\usepackage{mathrsfs}
\usepackage{nicefrac}
\usepackage{upgreek}
\usepackage{bbm}
\usepackage{psfrag}
\usepackage{hyperref}
\usepackage{cleveref}
\usepackage{hyperref}

\usepackage{enumitem}

\newcommand{\Rll}{R_{\ell \ell}}	
\usepackage{slashed}				

\begin{document}

\title{New Benchmark Models for Heavy Neutral Lepton Searches
{\bf \boldmath }}
\date{}
\author{Marco Drewes$^1$, Juraj Klaric$^1$, Jacobo Lopez Pavon$^2$\\ \\ 
	{\normalsize \it $^1$Centre for Cosmology, Particle Physics and Phenomenology,}\\ {\normalsize \it Universit\'{e} catholique de Louvain, Louvain-la-Neuve B-1348, Belgium}\\
	{\normalsize \it $^2$Instituto de F\'{\i}sica Corpuscular, Universidad de Valencia and CSIC}\\
	{\normalsize \it Edificio Institutos Investigaci\'on, Catedr\'atico Jos\'e Beltr\'an 2, 46980 Spain}\\
}
\maketitle
\thispagestyle{empty}
\begin{abstract}
	\noindent
	The sensitivity of direct searches for heavy neutral leptons (HNLs) in accelerator-based experiments depends strongly on the particles properties. Commonly used benchmark scenarios are important to ensure comparability and consistency between experimental searches, re-interpretations, and sensitivity studies at different facilities. In models where the HNLs are primarily produced and decay through the weak interaction, benchmarks are in particular defined by fixing relative strengths of their mixing with SM neutrinos of different flavours, and the interpretation of experimental data is known to strongly depend on those ratios. The commonly used benchmarks in which a single HNL flavour exclusively interacts with one Standard Model generation do not reflect what is found in realistic neutrino mass models. As a part of the activities within CERN's Physics Beyond Colliders initiative we identify two additional benchmarks, which we primarily select based on the requirement to provide a better approximation for the phenomenology of realistic neutrino mass models in view of present and future neutrino oscillation data.
\end{abstract}


\begin{small}
	\tableofcontents
\end{small}

\section{Introduction and summary}

Heavy Neutral Leptons (HNLs) that interact through weak interaction via their mixing with ordinary neutrinos (``neutrino portal") are a much studied extension of the Standard Model (SM) of particle physics. They represent one of the four portals to hidden sectors chosen by CERN's Physics Beyond Colliders (PBC) Working Groups \cite{Beacham:2019nyx,Agrawal:2021dbo} as well as case studies for future colliders \cite{FCC:2018evy,CEPCStudyGroup:2018ghi}. The physical motivation for HNLs is rich \cite{Abdullahi:2022jlv}; most notably they can be associated with the generation of light neutrino masses via the type-I seesaw mechanism ~\cite{Minkowski:1977sc,Glashow:1979nm,Gell-Mann:1979vob,Mohapatra:1979ia,Yanagida:1980xy,Schechter:1980gr}, can explain the observed baryon asymmetry asymmetry of the universe \cite{Canetti:2012zc} via leptogenesis \cite{Fukugita:1986hr}, and can serve as Dark Matter candidates \cite{Dodelson:1993je}.
Phenomenological studies are commonly based on the model 
\begin{equation}
 \mathcal L
\supset
- \frac{m_W}{v} \overline N \theta^*_\alpha \gamma^\mu e_{L \alpha} W^+_\mu
- \frac{m_Z}{\sqrt 2 v} \overline N \theta^*_\alpha \gamma^\mu \nu_{L \alpha} Z_\mu
- \frac{M}{v} \theta_\alpha h \overline{\nu_L}_\alpha N
+ \text{h.c.}
\ ,\label{PhenoModelLagrandian}
\end{equation}
were the spinor $N$ represents a single flavour of HNLs of mass $M$, 
while $\nu_{L \alpha}$ and $e_{L \alpha}$ are SM neutrinos and charged leptons, respectively, 
$h$ is the physical Higgs field and $v \simeq 174$ GeV is its vacuum expectation value. 
The HNLs' interaction strength is determined by the 
magnitudes $U_\alpha^2\equiv|\theta_\alpha|^2$ of their
mixing angles $\theta_\alpha$ with SM generation $\alpha = e,\mu,\tau$. Ignoring kinematical factors, the cross section for HNL production along with leptons of flavour $\alpha$ and their decay into gauge bosons and leptons of flavour $\alpha$ are both proportional to $U_\alpha^2$.\footnote{Many extensions of the SM predict HNLs with additional interactions, including new gauge interactions or an extended scalar sector.
In such models HNL production and decay may be governed by different parameters. 
For the present purpose we focus on scenarios in which all HNL interactions at the experimentally relevant energies can effectively be parametersied by their mixing with SM neutrinos.} 
$N$ can in principle be a Dirac or a Majorana spinor. 
In the latter case the ratio $\Rll$ between lepton number violating (LNV) and lepton number conserving (LNC) HNL decays is $\Rll=1$, while in the former case $\Rll=0$.

Realistic implementations of the type-I seesaw as the origin of neutrino masses necessarily require the existence of several HNL flavours; more precisely, the number $n$ of HNL flavours must be equal or larger than the number of massive SM neutrino, implying $n\geq 2$ if the lightest neutrinos mass $m_{\rm lightest}$ vanishes and $n\geq 3$ for $m_{\rm lightest} > 0$.
Depending on the HNL mass spectrum, their lifetime, and the underlying symmetries, the ratios observed at accelerator experiments may effectively interpolate between the cases $\Rll=0$ and $\Rll=1$ \cite{Deppisch:2015qwa,Anamiati:2016uxp}.  
Moreover, specific neutrino mass models make predictions for the HNL flavour mixing pattern, 
i.e., the relative size of the ratios $U_\alpha^2/U^2$ with 
$U^2=\sum_\alpha U_\alpha^2$, 
often presented in the form $U_e^2:U_\mu^2:U_\tau^2$. 
The sensitivity of experiments can strongly depend on these ratios even for fixed total mixing $U^2$ \cite{SHiP:2018xqw,Drewes:2018gkc,Tastet:2021ygq}. 

\begin{figure}
	\centering
	\includegraphics[width=0.5\textwidth]{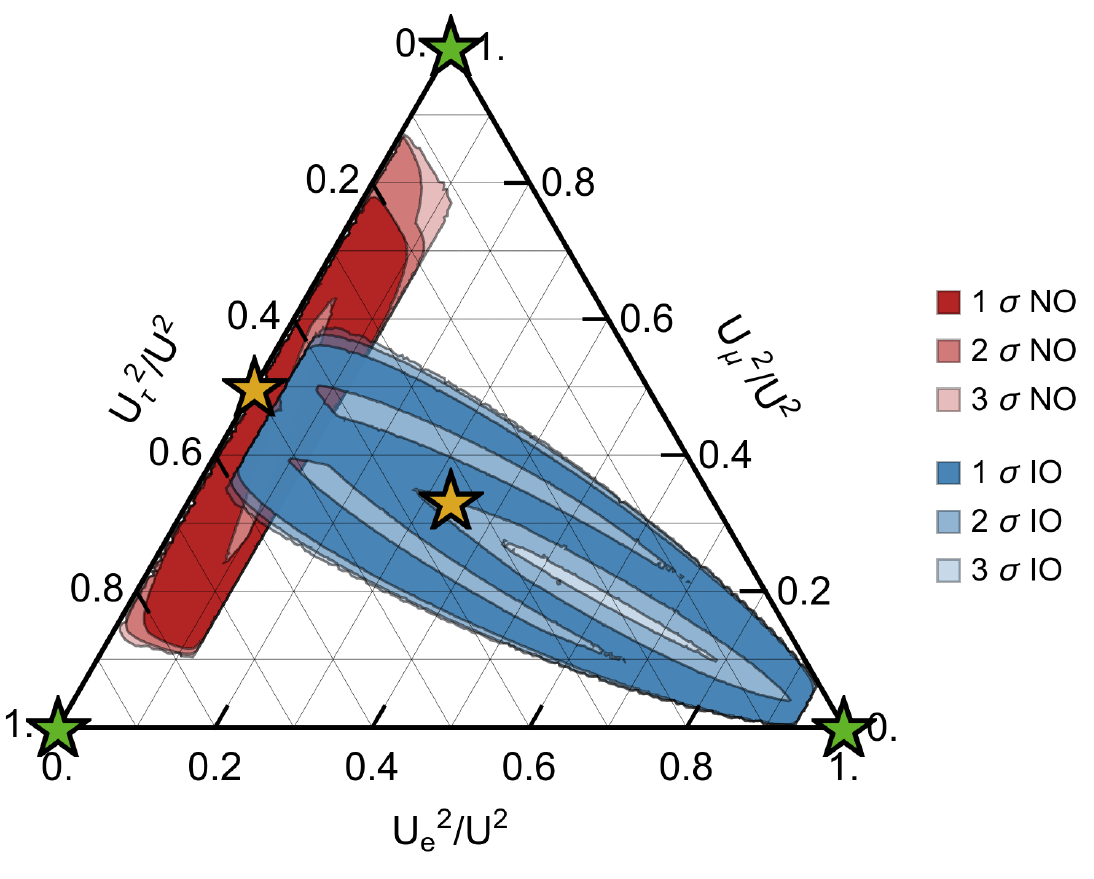}
	\caption{ The new benchmarks \eqref{NewBenchmarks} and old benchmarks \eqref{OldBenchmarks} for $U_e^2 : U_\mu^2 : U_\tau^2$ (here displayed in the form $U_\alpha^2/U^2$)
	 are represented by the yellow and green stars, respectively. They are compared to the range of $U_\alpha^2/U^2$ consistent with light neutrino oscillation data \cite{Esteban:2020cvm} for normal (red) and inverted (blue) ordering of light neutrino masses
		in the large mixing limit \eqref{LargeMixings} of the \emph{minimal seesaw model} \ref{ModelA}. In this limit the mixing angle ratios are completely fixed by the light neutrino oscillation parameters.
	}
	\label{fig:triangle2RHN}
\end{figure}

All of this gives rise to a rich phenomenology \cite{Atre:2009rg,Drewes:2013gca,Deppisch:2015qwa,Antusch:2016ejd,Chun:2017spz,Cai:2017mow,Abdullahi:2022jlv}.
Currently most searches assume that a single HNL flavour ($n=1$) couples exclusively to one SM generation (``single flavour mixing"), corresponding to the benchmarks BC6, BC7 and BC8 defined in \cite{Beacham:2019nyx},
\begin{subequations}\label{OldBenchmarks}
\begin{equation}\label{benchmarkB6}
    U_e^2 : U_\mu^2 : U_\tau^2 = 1 : 0 : 0 \quad \text{BC6}
    \end{equation}
    \begin{equation}
    U_e^2 : U_\mu^2 : U_\tau^2 = 0 : 1 : 0  \quad \text{BC7}\label{benchmarkB7}
    \end{equation}
    \begin{equation}
      U_e^2 : U_\mu^2 : U_\tau^2 = 0 : 0 : 1 \quad \text{BC8}\label{benchmarkB8}
      \end{equation}
\end{subequations}
with either $\Rll=1$ or $\Rll=0$.
While these provide a well-defined benchmark for experimental searches and sensitivity studies, they are inconsistent with the observed light neutrino properties. 
We propose an extended set of benchmark scenarios that can effectively describe many phenomenological aspects of realistic neutrino mass models within the Lagrangian \eqref{PhenoModelLagrandian}.  
Specifically, we propose the two new bencharks for the flavour mixing patterns,
\begin{subequations} \label{NewBenchmarks}
\begin{equation}
    U_e^2 : U_\mu^2 : U_\tau^2 = 0 : 1 : 1
    \label{NObenchmark}
    \end{equation}
    \begin{equation}
    U_e^2 : U_\mu^2 : U_\tau^2 = 1 : 1 : 1
    \label{IObenchmark}
    \end{equation}
\end{subequations}
while keeping the choices $\Rll=1$ and $\Rll=0$.
The new and previous benchmarks \eqref{NewBenchmarks} and \eqref{OldBenchmarks}, respectively, are compared in figure \ref{fig:triangle2RHN}.
The remainder of this article is organised as follows. In section \ref{sec:SelectionCriteria} we outline the criteria on which we base the benchmark selection. In section \ref{sec:MinimalModel} we apply these criteria to motivate the proposed benchmark in the minimal type-I seesaw model. In section \ref{sec:ThreeHNLModel} we comment on the robustness of the motivation for the new benchmarks when going beyond that minimal model.  

The study presented here was prepared for the PBC Working Group Meeting in December 2021 as recommendation for the experimental collaborations within PBC. However, the physical motivation for choosing the new benchmarks are based on the requirement to capture the HNL properties predicted by realistic neutrino mass models (and not on the properties of specific PBC experiments), they may also be used for searches, reinterpretations and sensitivity projections of collider experiments.
The motivation in using multiple benchmark scenarios is based on the observation that the sensitivity of both fixed target experiments \cite{SHiP:2018xqw,Drewes:2018gkc} and LHC experiments \cite{Tastet:2021vwp,ATLAS:2022atq} can change by orders of magnitude when e.g.~changing the flavour mixing pattern.

\section{Criteria for the benchmark selection}\label{sec:SelectionCriteria}

For given HNL mass $M$ a benchmark in the sense considered here is defined by fixing the ratios $U_e^2:U_\mu^2:U_\tau^2$ as well as $\Rll$. 
For the sake of definiteness, we base the benchmark selection on the requirement to explain the known properties of the light neutrino mass and mixing matrices $m_\nu$ and $V_\nu$ in models that can effectively be described within the pure type-I seesaw Lagrangian, i.e., the extension of the SM by $n$ flavours of right-handed neutrinos,
and apply the following criteria.
\begin{enumerate}[label = \arabic*)]
\item\label{eins} \textbf{Consistency with neutrino oscillation data}.
Our key requirement is to effectively reproduce the phenomenology of realistic neutrino mass models within the phenomenological model \eqref{PhenoModelLagrandian}. 
In comparison to previous studies that took a similar approach \cite{SHiP:2018xqw,Drewes:2018gkc,Tastet:2021ygq} we use an updated global fit to light neutrino oscillation data to constrain the $U_e^2:U_\mu^2:U_\tau^2$, and we take into consideration how those global fits can evolve when the Dirac phase $\delta$ is measured in future experiments. The latter ensures a maximal robustness of the selected benchmarks with respect to future neutrino oscillation data.
\item\label{zwei} \textbf{Added value}. In view of the considerable effort that any experimental search requires, new benchmark scenarios can only be justified if they lead to phenomenological predictions for accelerator-based experiments that are clearly distinguishable from the existing benchmarks B6-B8. Past studies have shown that percent-level changes in the $U_e^2:U_\mu^2:U_\tau^2$ only affect the experimental sensitivities near the single-flavour benchmarks B6-B8 in \eqref{OldBenchmarks}, in particular near the B8 \cite{Drewes:2018gkc}.  
\item\label{drei} \textbf{Symmetry considerations}. We generally follow an agnostic approach that makes \ref{eins} and \ref{zwei} the most important criteria. If these criteria do not conclusively prefer one amongst several possible choices of the $U_e^2:U_\mu^2:U_\tau^2$, we take into account which one of them can be motivated by model building considerations, in particular discrete symmetries of the fermion mixing matrices (cf.~\cite{King:2017guk,Xing:2020ijf}).
In particular, we consider maximal CP-violation in the direction of the current experimental hints appealing.
\item\label{vier} \textbf{Simplicity.} Between benchmarks that are similarly strongly motivated by the criteria \ref{eins}-\ref{drei}, we give preference to ratios $U_e^2:U_\mu^2:U_\tau^2$ that are simple and can be easily communicated to the community. 
\item\label{fuenf} \textbf{Leptogenesis}. A second motivation for the existence HNLs with sub-TeV masses (in addition to explaining the light neutrino masses) lies in the possibility to explain the observed baryon asymmetry of the universe through low scale leptogenesis \cite{Akhmedov:1998qx,Pilaftsis:2003gt,Asaka:2005pn}. Amongst those candidates for benchmark models that are similarly well-motivated by the criteria \ref{eins}-\ref{vier}, scenarios that can accommodate leptogenesis in an experimentally accessible range of $M$ and $U^2$ are favoured.  
\end{enumerate}

\section{Benchmark selection}

We consider two types of scenarios, both based on the most general renormalisable extension of the SM by $n$ flavours of right-handed neutrinos and no other new particles,
\begin{enumerate}[label=\Alph*)]
\item\label{ModelA} \textbf{Minimal seesaw model}.
The minimal realisation of the type-I seesaw Lagrangian that is consistent with all light neutrino oscillation data includes two HNLs ($n=2$), which implies $m_{\rm lightest}=0$ up to one-loop level \cite{Davidson:2006tg}. 
\item\label{ModelB} \textbf{Next-to-minimal seesaw model}. The minimal realisation of the type-I seesaw Lagrangian that can give mass to all three SM neutrinos ($m_{\rm lightest}>0$) includes three HNLs ($n=3$). This choice is also necessary for anomaly cancellation in any extension of the SM that includes a gauged $U(1)_{B-L}$ symmetry, where $B$ and $L$ refer to the baryon and lepton number, respectively.
\end{enumerate}

\subsection{The minimal model \ref{ModelA}}\label{sec:MinimalModel}
From the viewpoint of direct searches for HNLs at accelerator based experiments, the interesting scenarios are those with $M$ below the TeV scale and mixing angles that are much larger than the naive seesaw expectation,
   \begin{equation}\label{LargeMixings}
    \begin{tabular}{c}
        $U^2 \gg \sqrt{\sum_i m_i^2}/M$
        \end{tabular}
    \end{equation}
A consistent theoretical framework in which  these conditions are fulfilled is represented by the class of
neutrino mass models that can effectively be described within 
the minimal seesaw model \ref{ModelA} in the
\emph{symmetry protected scenario} in which the two right-handed neutrinos approximately respect a generalisation of the global $U(1)_{B-L}$ symmetry known in the SM \cite{Shaposhnikov:2006nn,Kersten:2007vk}. This implies that the two HNLs $N_I$ 
(with $I=1,2$) 
have quasi-degenerate masses, and their mixing angles $\theta_{\alpha I}$ with individual SM flavours approximately fulfil the relation $\theta_{\alpha 2} \simeq i \theta_{\alpha 1}$.
This has several advantages.
\begin{itemize}
    \item The presence of a generalised $B-L$ symmetry that suppresses the light neutrino masses $m_i$ is a necessary condition to make mixing angles \eqref{LargeMixings}
    that are large enough to detect the HNLs at in accelerator-based experiments 
    consistent with the smallness of the $m_i$ without fine-tuning \cite{Kersten:2007vk,Moffat:2017feq}. It permits technically natural low scale seesaw models with $M$ below the electroweak scale and coupling constants of order one, cf.~section 5.1 in \cite{Agrawal:2021dbo}.
    \item This model effectively describes the phenomenology of popular extensions of the SM, including the Neutrino Minimal Standard Model ($\nu$MSM)~\cite{Asaka:2005pn,Asaka:2005an} and inverse \cite{Wyler:1982dd,Mohapatra:1986aw,Mohapatra:1986bd,Bernabeu:1987gr,Branco:1988ex} and linear~\cite{Akhmedov:1995ip,Akhmedov:1995vm,Gavela:2009cd} seesaws scenarios.
    \item Its minimality makes the model highly predictive \cite{Hernandez:2016kel,Drewes:2016jae}. In particular, in the phenomenologically relevant regime \eqref{LargeMixings} the ratios $U_e^2:U_\mu^2:U_\tau^2$ are in good approximation determined by the parameters in the light neutrino mixing matrix $V_\nu$ alone.
    \item The model is self-consistent in the sense that it could in principle be a complete effective field theory up to the Planck scale \cite{Bezrukov:2012sa}.
    \item The model allows for low scale leptogensis in the range of $M$ and $U_\alpha^2$ that is accessible to experiments, cf.~\cite{Klaric:2020phc,Klaric:2021cpi,Hernandez:2022ivz} for an updated parameter space scan.
\end{itemize}
Based on the criteria \ref{eins}-\ref{fuenf} we propose the two new benchmarks for the flavour mixing pattern given in \eqref{NewBenchmarks}. 
The new benchmarks \eqref{NewBenchmarks} are compared to the existing ones \eqref{OldBenchmarks} in figure \ref{fig:triangle2RHN}. 
If relevant for the respective search, we further propose considering the choices $\Rll=1$ and $\Rll=0$ for each benchmark in \eqref{OldBenchmarks} and \eqref{NewBenchmarks}.

\paragraph{Dirac vs Majorana HNLs.}
The choices $\Rll=1$ and $\Rll=0$ correspond to assuming a pure Majorana or a pure Dirac HNL in the phenomenological model \eqref{PhenoModelLagrandian}. In technically natural realisations of the minimal realistic seesaw model \ref{ModelA}, $\Rll$ can take any value between $0$ and $1$ \cite{Drewes:2019byd},\footnote{Within the reach of near-future experiments, the requirement to avoid tunings points towards values $\Rll\sim 1$ for $M\ll m_W$ and $\Rll\ll 1$ for $M\gg m_W$ in the model \ref{ModelA}, with a sizeable regime in between in which any value $0\leq\Rll\leq1$ can be realised.}
An experimental determination of $\Rll$ is in principle possible through several observables 
(including direct observations of the ratio between lepton number violating and conserving decays \cite{Anamiati:2016uxp,Das:2017hmg,Abada:2019bac},
its dependence on position \cite{Cvetic:2015ura,Antusch:2017ebe,Cvetic:2018elt}, the 
momentum distribution \cite{Dib:2017iva,Arbelaez:2017zqq,Balantekin:2018ukw,Hernandez:2018cgc,Tastet:2019nqj,Blondel:2021mss,Blondel:2021mss} and polarisation \cite{Blondel:2021mss} of the decay products, 
the $U_\alpha^2/U^2$ \cite{Dib:2016wge}, and the lifetime \cite{Alimena:2022hfr}), and would be very interesting from the viewpoint of testing the underlying model \cite{Hernandez:2016kel,Drewes:2016jae,Antusch:2017pkq}.
However, performing analyses for a sizeable number of values for $\Rll$ would be a considerable effort, and an additional difficulty is posed by the fact that current event generators cannot simulate the HNL oscillations that can occur for values that interpolate between $\Rll=0$ and $\Rll=1$ \cite{Boyanovsky:2014una,Cvetic:2015ura,Antusch:2017ebe,Cvetic:2018elt,Antusch:2020pnn} in a straightforward way.
Due to these technical and practical limitations, at this stage it seems reasonable to continue using the limiting cases  $\Rll=0$ and $\Rll=1$ for which software tools are readily available (cf.~e.g.~\cite{Alva:2014gxa,Degrande:2016aje,Pascoli:2018heg,Coloma:2020lgy}), knowing that realistic neutrino mass models are likely to interpolate between these cases.

\paragraph{The flavour mixing pattern.}
We apply the criteria \ref{eins}-\ref{fuenf} to the minimal seesaw model \ref{ModelA} in order to motivate the new benchmarks \eqref{NewBenchmarks}.
Our primary selection criterion is \ref{eins}, i.e., the consistency with current and future neutrino oscillation data. 
Due to its minimality, the model \ref{ModelA} can make testable predictions for the ratios $U_\alpha^2/U^2$, cf.~figure \ref{fig:triangle2RHN}.
The flavoured mixing ratios $U_e^2:U_\mu^2:U_\tau^2$ are determined 
by the parameters in the light neutrino mass and mixing matrices $m_\nu$ and $V_\nu$ alone \cite{Hernandez:2016kel,Drewes:2016jae}, with the main uncertainties currently coming from the lack of knowledge about the CP phase $\delta$ and the Majorana phase $\alpha$\footnote{Note that there is only one physical Majorana phase in $V_\nu$ for $n=2$.}.
This can be nicely displayed in the form of ternary diagrams \cite{Caputo:2017pit}, cf.~figure Fig.~\ref{fig:triangle2RHN}.
As we are mapping the seven neutrino oscillation parameters (three angles, two mass splittings and two phases) onto a two-dimensional space of flavour ratios, each choice of ratios $U_\alpha^2/U^2$ corresponds to a continuum of neutrino oscillation parameters.
Using the Casas-Ibarra parameterisation \cite{Casas:2001sr}, we vary the neutrino oscillation parameters within the $3 \sigma$ allowed range from NuFIT 5.1~\cite{Esteban:2020cvm}, and we keep the Majorana phase as a free parameter $\in [0, 4 \pi]$~\cite{Molinaro:2008rg}.
For each choice of parameters one can estimate the $\chi^2$ by combining the 1-d projections for $\Delta m_{ij}^2$, $s_{12}$ and $s_{13}$ with the 2-d $\chi^2$ projection for the parameters $s_{23}$ and $\delta$, which have the largest deviation from Gaussianity.
We follow the procedure from~\cite{Bondarenko:2021cpc} and collect the points generated this way into $U_e^2:U_\mu^2:U_\tau^2$ bins and take the minimal value of $\chi^2$.\footnote{Alternatively, one can instead assign a posterior probability to the flavour ratios as was done in~\cite{Drewes:2018gkc}.} 
The results are shown in Fig.~\ref{fig:triangle2RHN}.
There is a significant degeneracy among the allowed mixing angles, as the ratios $U_\alpha^2/U^2$ have a strong  dependence on the unknown Majorana phase.
Besides this 1-dimensional degeneracy, the strongest dependence remains on the CP phase $\delta$, and the value of $s_{23}$ as the remaining parameters are already measured at a much higher precision.
In practice it is therefore sufficient to only vary the Majorana phase, and $s_{23}^2$ and $\delta$ in the allowed $3 \sigma$ regions, and fix the remaining parameters to their best fit values.

\begin{figure}
	\centering
    \includegraphics[width=0.5\textwidth]{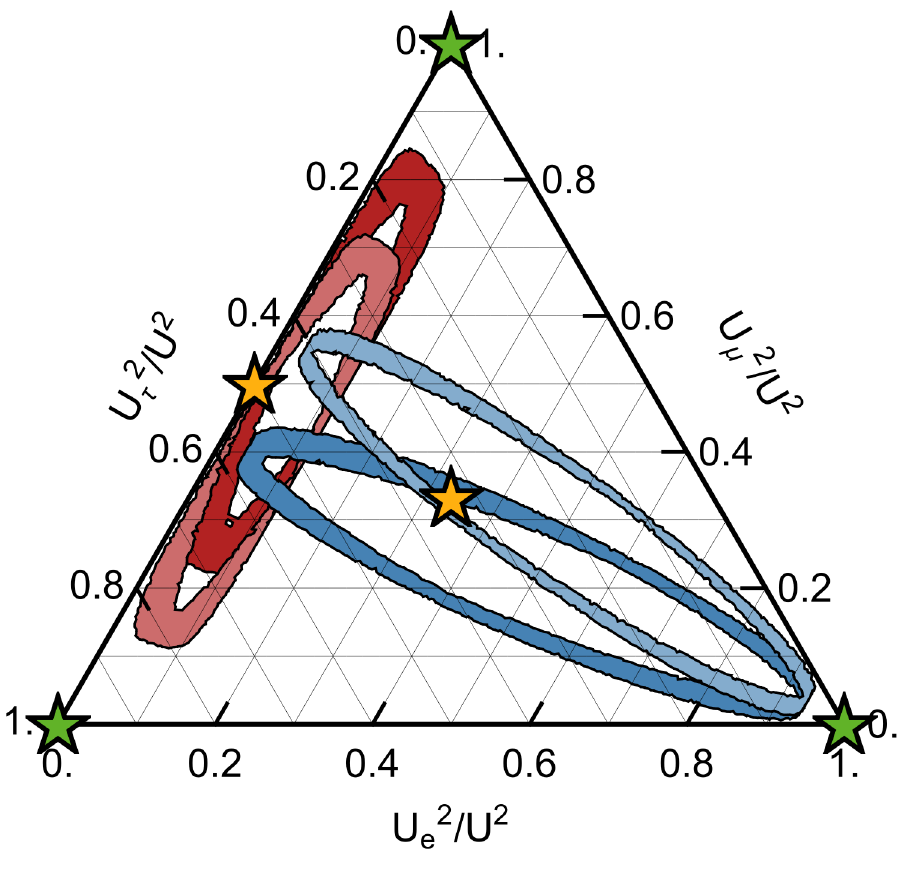}
	\caption{
		Projected $90\%$ CL for the allowed range of $U_\alpha^2/U^2$ after 15 years of data taking at DUNE \cite{DUNE:2020jqi} in case of normal ordering (red) and inverted ordering (blue).
For this forecast we used the $\chi^2$ projections from 
\cite{DUNE:2020ypp} that were generated assuming the true values $\delta=-\pi/2$ and two benchmark values of the light neutrino mixing angle $s_{23}^2\equiv \sin^2\theta_{23}=0.58$ (darker region) and $s_{23}^2=0.42$ (lighter region) that were used in the DUNE TDR \cite{DUNE:2020ypp}. 
To generate these contours we use a random sample of points within the expected sensitivity bounds for the parameters $\delta$ and $s_{23}$, while the remaining parameters are fixed to their best fit values.
	}
	\label{fig:triangle2RHNDUNE}
\end{figure}

\begin{figure}
	\centering
	\includegraphics[height=0.3\textheight]{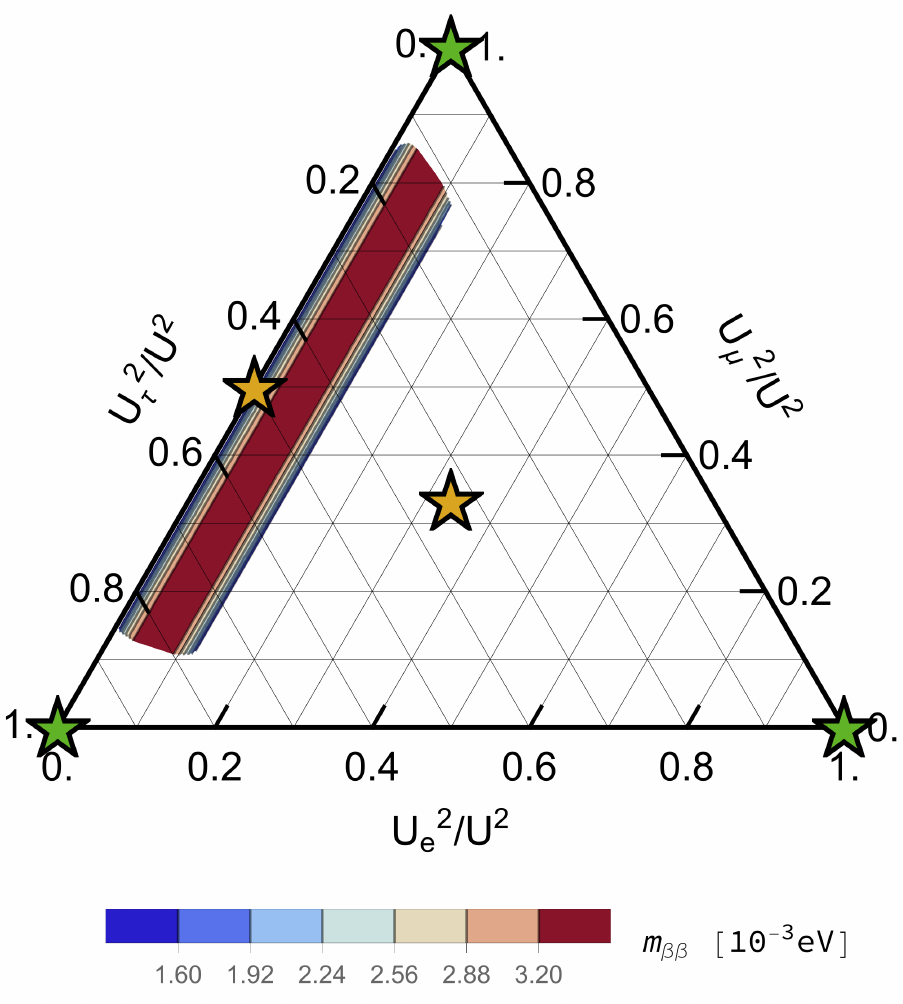} \includegraphics[height=0.3\textheight]{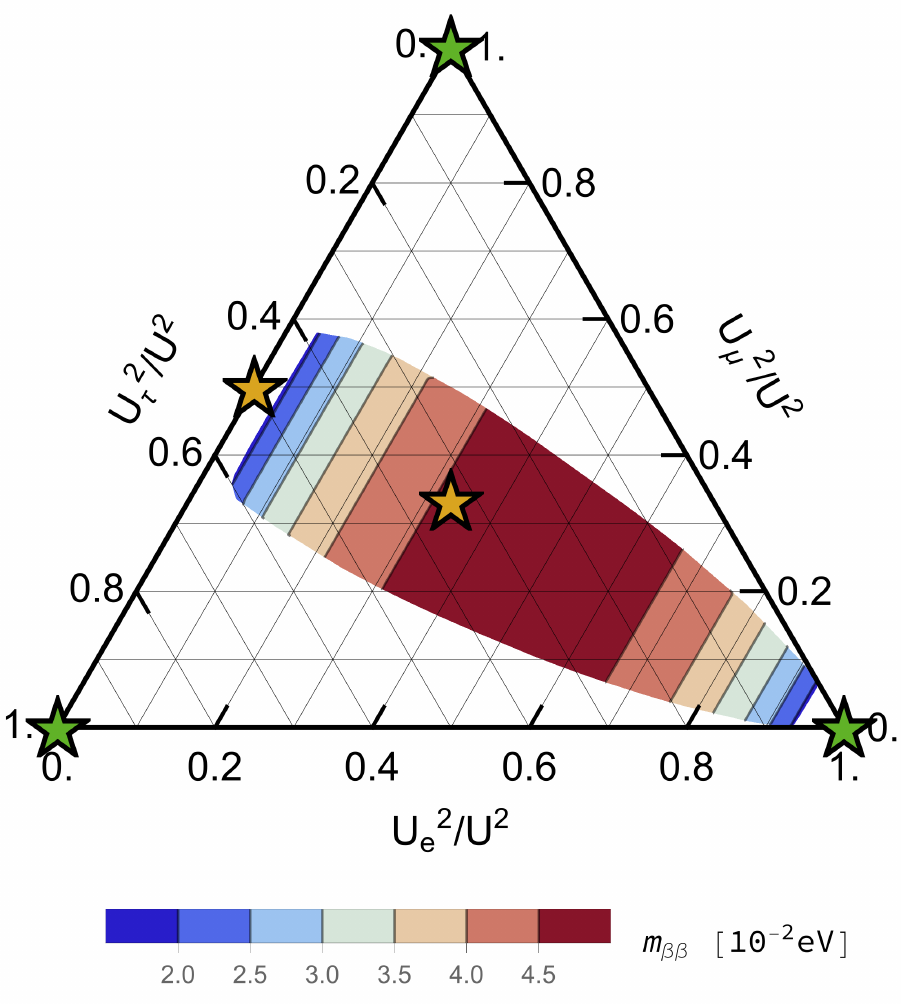}
	\caption{
		The neutrinoless double $\beta$-decay parameter $m_{\beta\beta}$ in model \ref{ModelA}
		as a function of the flavour mixing ratios $U_\alpha^2/U^2$ in the case of normal ordering (left) and inverted ordering (right) in the limit \eqref{LargeMixings} and neglecting the contribution from HNL exchange. 
		Due to the significant uncertainties in the nuclear matrix elements, the value of $m_{\beta\beta}$ cannot be measured precisely enough to determine the Majorana phase.
		Nonetheless, it can still impose constraints on the allowed values of $U_\alpha^2/U^2$.
		Finally, we note that the  benchmark point \eqref{IObenchmark} lies exactly in the region where $m_{\beta \beta}$ is close to its maximal value for inverted ordering.
	}
	\label{fig:triangle2RHN0nubb}
\end{figure}

\begin{figure}
	\centering
	\includegraphics[height=0.22\textheight]{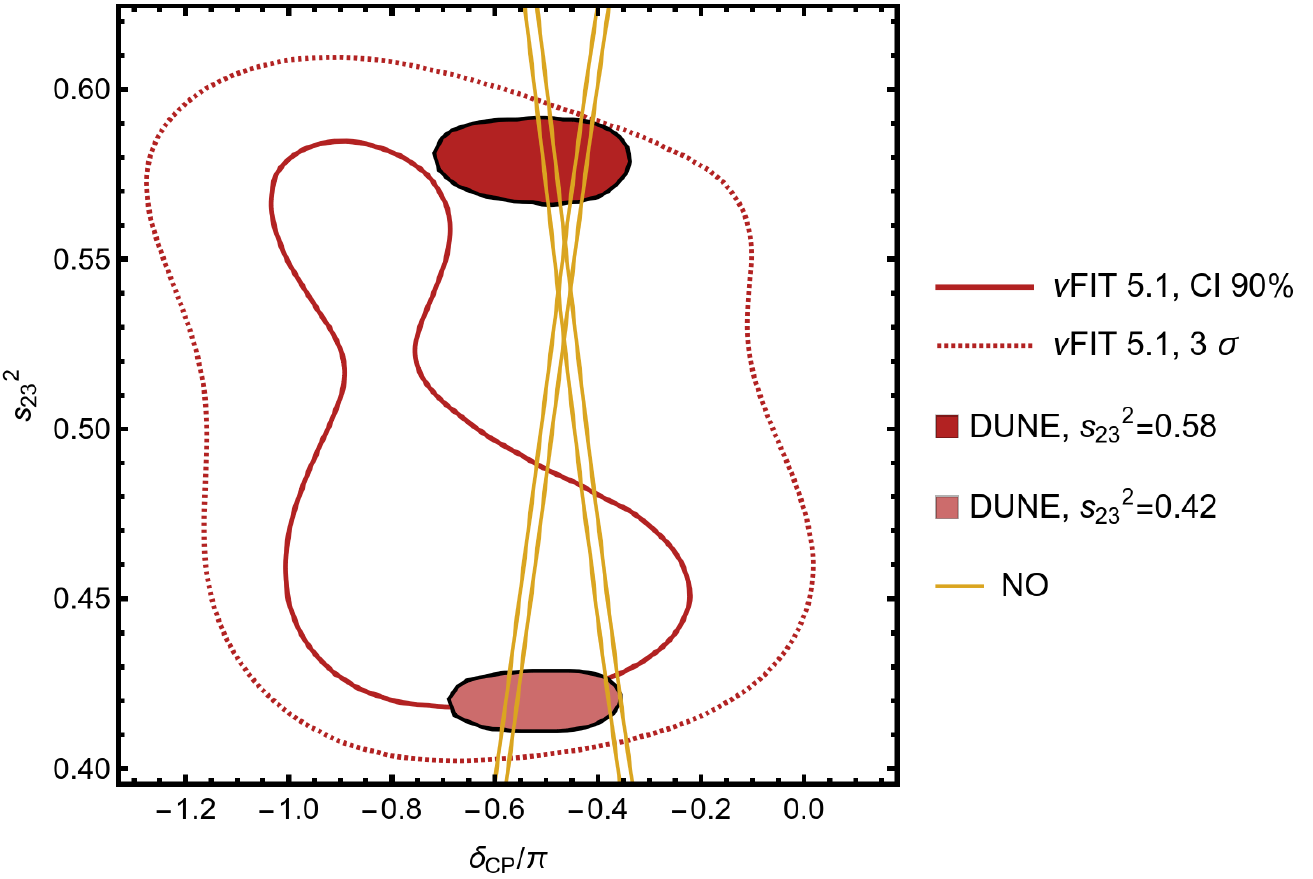}
    \includegraphics[height=0.22\textheight]{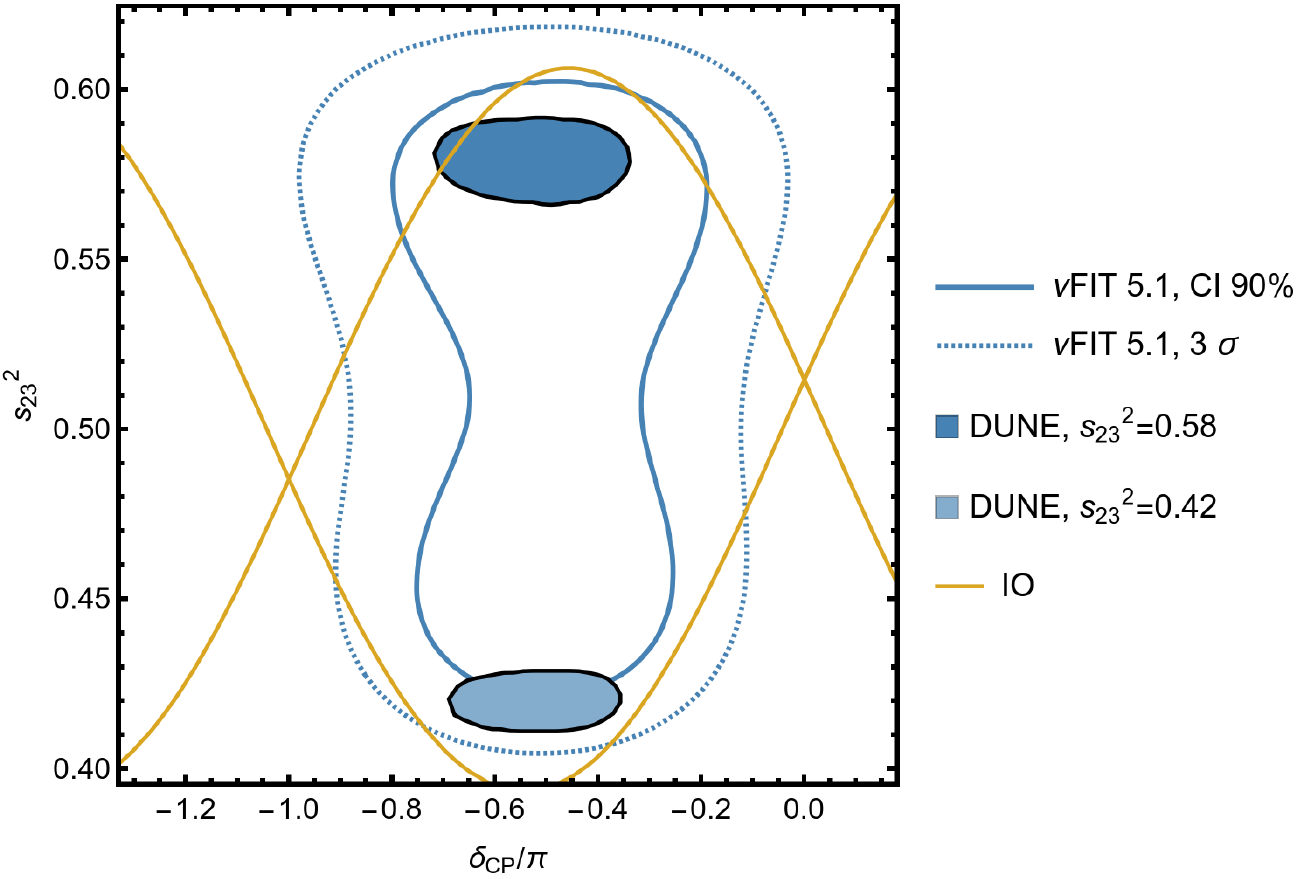}
	\caption{The regions inside the lines represent the regions in the $s_{23}^2$ and $\delta$  plane preferred by the global fit NuFit 5.1 to light neutrino oscillation data for normal ordering (red, left panel) and inverted ordering (blue, right panel). The dotted line corresponds to $3 \sigma$ CL, the full one to $90\%$ CL. The filled regions are projections for DUNE data, using the values and colour coding as in figure \ref{fig:triangle2RHNDUNE}. The yellow lines indicate combinations of $s_{23}^2$ and $\delta$ that correspond to the combinations \eqref{AlmostBenchmark} (left panel) and \eqref{IObenchmark} (right panel), i.e., along the lines the ratios $U_\alpha^2/U^2$ are constant. 
	}
	\label{fig:DeltaDependence}
\end{figure}

An interesting observation is that the centres of the allowed regions are already slightly disfavoured by current neutrino oscillation data, meaning that the naive approach of placing a benchmark somewhere in the middle of these region is not the best way 
to ensure robustness with respect to future neutrino oscillation data.
This is a result of the current information about $\delta$. Once $\delta$ is measured, the allowed region will further reduce,  and a measurement of the $U_\alpha^2/U^2$ would allow to measure the Majorana phase \cite{Hernandez:2016kel,Drewes:2016jae,Caputo:2016ojx}, provided that $\delta$ is measured independently at DUNE \cite{DUNE:2020jqi} or HyperK \cite{Hyper-Kamiokande:2018ofw}.
This would in turn permit to predict the rate of neutrinoless double $\beta$-decay in the model \ref{ModelA}, cf.~figure \ref{fig:triangle2RHN0nubb}.
We show in figure \ref{fig:triangle2RHNDUNE} how the allowed range of $U_\alpha^2/U^2$ is expected to change with future neutrino oscillation experiments, using DUNE as an example. 
A comparison between the allowed regions for the two chosen values of $s_{23}^2$ in that figure suggests that benchmarks with 
\begin{eqnarray}\label{NO:roughly}
U_e^2\ll U_\mu^2\simeq U_\tau^2 
\end{eqnarray}
are the ones that are most safe from being ruled out in the case of normal ordering (the allowed region roughly moves up and down parallel to the $U_e^2=0$ line in figure \ref{fig:triangle2RHNDUNE} when changing $s_{23}^2$), while scenarios with
\begin{eqnarray}\label{IO:roughly}
U_e^2\gg U_\mu^2\simeq U_\tau^2
\end{eqnarray}
are most the most safe ones in case of inverted ordering (the allowed region roughly rotates around the benchmark point \eqref{benchmarkB7} in the $U_\mu^2$-corner when changing $s_{23}^2$). However, points of the type \eqref{IO:roughly} are disfavoured by criterion \ref{zwei}: They would provide little added value, as they are relatively close to the existing benchmark \eqref{benchmarkB6}.
At the same time inverted ordering clearly allows for points in which all three mixings are of comparable size,
\begin{eqnarray}\label{FlavourCommunism}
U_e^2 \simeq U_\mu^2 \simeq U_\tau^2,
\end{eqnarray}
a scenario that is not reflected in the current benchmarks \eqref{OldBenchmarks}. 
The criterion \ref{zwei} implies that we should add a benchmark of the kind \eqref{FlavourCommunism} to capture the qualitatively different phenomenology that one can expect from this compared to the benchmarks \eqref{OldBenchmarks}. Neither current neutrino oscillation data in figure \ref{fig:triangle2RHN} nor the projections in figure \eqref{fig:triangle2RHNDUNE} single out a particular point amongst all combinations of the type \eqref{FlavourCommunism} (keeping in mind that any value between the two chosen examples for $s_{23}^2$ is possible). Since all combinations of the type \eqref{FlavourCommunism} are expected to lead to similar phenomenology at accelerator experiments, the simplicity criterion \ref{vier} can be used to argue for the equipartition benchmark \eqref{IObenchmark} to represent them. The benchmark \eqref{IObenchmark} can in addition be motivated by criterion \ref{drei}: As illustrated in figure \ref{fig:DeltaDependence}, the statistically favoured region is roughly centred around the value $\delta=-\pi/2$.
Finally, benchmark \eqref{IObenchmark} can be used when connecting 
accelerator-based experiments to neutrinoless double $\beta$-decay.
In the large mixing limit \eqref{LargeMixings} and neglecting the HNL contribution to $m_{\beta \beta}$ (which is justified for sufficiently mass-degenerate HNLs 
		and $M \gg  160$ MeV \cite{Bezrukov:2005mx}\footnote{Note that the contribution from HNLs can in principle be sizeable \cite{Blennow:2010th,Lopez-Pavon:2012yda} and play a crucial role for the testability of the model \ref{ModelA} and leptogenesis \cite{Hernandez:2016kel,Drewes:2016lqo}.
		}), knowledge of $U_e^2/U^2$ determines $m_{\beta\beta}$ in the symmetry protected limit \eqref{LargeMixings} of model \ref{ModelA}:
		\begin{align}
		    |m_{\beta \beta}|^2  =
		    \begin{cases}
		            U_e^2/U^2 (m_2+m_3) [m_2 (2 s_{12}^2 c_{13}^2 - U_e^2/U^2) + m_3 (2 s_{13}^2 - U_e^2/U^2)] & \text{for NO, and,}\\
		            U_e^2/U^2 (m_1+m_2) [m_1 (2 c_{12}^2 c_{13}^2 - U_e^2/U^2) - m_2 (2 s_{12}^2 s_{13}^2 + U_e^2/U^2)] & \text{for IO,}
		    \end{cases}
		\end{align}
		as shown in figure~\ref{fig:triangle2RHN0nubb}.
		The benchmark \eqref{IObenchmark} lies in the region where $m_{\beta \beta}$ is close to its maximal value for inverted ordering.
		Finally, it is worthwhile mentioning that leptogenesis in the model \ref{ModelA} is feasible for this benchmark \cite{Hernandez:2016kel,Drewes:2016jae,Antusch:2017pkq,Hernandez:2022ivz}.

Turning to the case of normal ordering, we notice that literally all allowed points exhibit the hierarcy $U_e^2\ll U_\mu^2,U_\tau^2$. Amongst these, the criteria \ref{eins} and \ref{zwei} both prefer points of the kind \eqref{NO:roughly}: Those are the most robust against varying $s_{23}$ in figure \ref{fig:triangle2RHNDUNE}, and they are the furthest away from the existing benchmarks \eqref{OldBenchmarks}, potentially giving rise to qualitatively different phenomenology in direct searches for HNLs. If we fix $U_\mu^2 \approx U_\tau^2$, the points with $U_e^2/U^2\sim 0.005$ and $U_e^2/U^2 \sim 0.12$ are favoured from the viewpoint of figure \ref{fig:triangle2RHNDUNE}. 
Between them the smaller one, which leads to the combination 
\begin{equation}\label{AlmostBenchmark}
U_e^2 : U_\mu^2 : U_\tau^2 = 2 : 199 : 199
\end{equation}
is favourable with respect to criterion \ref{zwei}, as it is further away from \eqref{IObenchmark}.
Further, being consistent with values $\delta=-\pi/2$ (cf.~figure \ref{fig:DeltaDependence}), it obeys criterion \ref{drei}.
The combination \eqref{AlmostBenchmark} has the disadvantage that it involves 
fractions, 
which is in tension with 
with criterion \ref{vier}. These shortcomings can be overcome by using \eqref{NObenchmark} instead of \ref{AlmostBenchmark},  
which comes at the price that \eqref{NObenchmark} strictly speaking cannot reproduce the observed light neutrino oscillation data in the symmetry protected model \ref{ModelA} (as it would lead to $\nu_{L e}$ not mixing with other flavours). However, for all observables that do not crucially rely on a non-zero $U_e^2$, the phenomenology of the combinations \eqref{AlmostBenchmark} and \eqref{NObenchmark} is likely to be very similar.
This applies to most direct searches at experiments involving proton beams. 
At $e^+ e^-$-colliders \eqref{NObenchmark} is preferred from the viewpoint of the added value criterion \ref{zwei} because it suppresses charged current HNL production at tree level, hence comparing  \eqref{NObenchmark} to \eqref{IObenchmark} as well as \eqref{OldBenchmarks} permits to map out the most optimistic and most pessimistic scenarios.\footnote{These considerations are tailored to direct searches for HNLs.
In some important indirect probes (such as neutrinoless double $\beta$-decay, $\mu\to e \gamma$ decays, and $\mu$ conversion in nuclei) the small but non-zero $U_e^2$ required by neutrino oscillation data is crucial, and a benchmark as \eqref{AlmostBenchmark} can be useful. The same caveat holds for direct searches involving triggers or vetos that rely on charged lepton flavour violating processes including electrons.
This e.g.~explains the surprisingly large difference observed between benchmarks 5 and 6 defined in \cite{Tastet:2021vwp} (which closely resemble our new benchmarks) for the case of Dirac HNLs ($\Rll=0$), which is caused by the opposite-sign same-flavour veto.
}
Finally, the benchmark \eqref{NObenchmark} is well-motivated from the viewpoint of criterion \ref{fuenf}, as leptogenesis for the largest $U^2$ in the model \ref{ModelA} requires a strong hierarcy between the $U_\alpha^2/U^2$. In particular, for normal ordering configurations of the kind \eqref{NO:roughly} allow for the largest $U^2$ consistent with leptogenesis \cite{Hernandez:2016kel,Drewes:2016jae,Antusch:2017pkq}, and hence the largest production cross section in particle collisions.

\subsection{The model \ref{ModelB} with three HNLs}\label{sec:ThreeHNLModel}
The type-I seesaw model with three HNLs can give masses to all three SM neutrinos. This comes at the price of a considerably larger parameter space comprising 18 instead of the 11 new parameters in the  model \ref{ModelA} with two HNLs. 
As a result, the model is far less predictive, making it more difficult to single out well-motivated benchmarks. 
The $U_\alpha^2/U^2$ are not determined by light neutrino parameters in $m_\nu$ and $V_\nu$ and more, but also depend on phases in the HNL sector. As a result, knowledge of the $U_\alpha^2/U^2$ and light neutrino parameters is not sufficient any more to compute $m_{\beta\beta}$.  The ranges of values for the $U_\alpha^2/U^2$ that are consistent with neutrino oscillation data increases with $m_{\rm lightest}$. For values $m_{\rm lightest}< 10^{-5}$ eV they roughly agree with figure \ref{fig:triangle2RHN}, but already for $m_{\rm lightest} \sim 10^{-2}$ eV they cover almost the entire triangle for both light neutrino mass orderings \cite{Chrzaszcz:2019inj}.
Hence, neither laboratory measurements nor cosmology will impose a sufficiently strong upper bound on $m_{\rm lightest}$ in the near future to substantially restrict the range of $U_\alpha^2/U^2$.  
The maximal $U^2$ consistent with leptogenesis is several orders of magnitude larger than in the minimal model \ref{ModelA} with two HNLs, but there is no clear preference for hierarchical $U_\alpha^2/U^2$ because the large $U^2$ can be achieved through other dynamical effects \cite{Abada:2018oly,Drewes:2021nqr}.
As a result the range of $U_\alpha^2/U^2$ can be treated as unconstrained for all practical purposes. 
Within specific models that exhibit additional symmetries (such as discrete flavour symmetries) this range can be restricted, but such restrictions are strongly model dependent. If any data in the future points to a particular class of models, or if a consensus in the scientific community is reached to identify a particular set of symmetries as reference models, this would permit to define additional benchmarks for the $U_e^2:U_\mu^2:U_\tau^2$.  
Within the agnostic approach we take, and in view of the present data, there appears to be no strong motivation to add any further benchmarks to \eqref{OldBenchmarks} and \eqref{NewBenchmarks}.
 
 
 
\section{Conslusions}

The interpretation and re-interpretation of experimental searches for HNLs are greatly facilitated when different experiments perform their analyses for a commonly used set of benchmark scenarios. Using the same set of benchmarks also allows to make fair comparisons of the sensitivity of proposed future experiments, aiding institutions and funding agencies in their decisions between different proposals. 
In the context of HNLs that exclusively interact through the SM weak interaction in \eqref{PhenoModelLagrandian} at the energies relevant for a given experiment, a benchmark comprises a selection of ratios $U_e^2:U_\mu^2:U_\tau^2$ (with $U_\alpha^2=|\theta_\alpha|^2$), along with a specification of the ratio $\Rll$ between LNV and LNC HNL decays.
In \cite{Beacham:2019nyx} the PBC community identified the three benchmarks \eqref{OldBenchmarks}, which have been used by both PBC and collider experiments.
While these single-flavour benchmarks provide well-defined reference points, a shortcoming is that they do not describe the properties of HNLs predicted by realistic neutrino mass models well. In the present article we propose two additional benchmarks \eqref{NewBenchmarks} to alleviate this shortcoming.
They are selected based on the criteria \ref{eins}-\ref{fuenf} defined in section \ref{sec:SelectionCriteria}, with the main goal to provide a better approximation for the phenomenology of realistic neutrino mass models within the framework of the phenomenological Lagrangian \eqref{PhenoModelLagrandian}. The old and new benchmarks are shown in figure \ref{fig:triangle2RHN}. In combination they provide an extended coverage of the range of possible HNL properties, and in particular effectively approximate the phenomenology of neutrino mass models within the type-I seesaw framework in PBC as well as collider experiments.

The combined sets \eqref{OldBenchmarks} and \eqref{NewBenchmarks} clearly cannot capture all phenomenological aspects of complete neutrino mass models, not only because they approximate the continuous ratios $U_e^2:U_\mu^2:U_\tau^2$ by a discrete set, but also because there are phenomena that can qualitatively not be described within the single HNL Lagrangian \eqref{PhenoModelLagrandian}, such as HNL oscillations in the detector and related effects, including CP-violation and non-integer values of $\Rll$ that can depend on the HNL boost and displacement. In case any HNLs are discovered experimentally, all these phenomena will provide important keys to understand their properties, what role they may play for the mechanism of neutrino mass generation, baryogenesis, dark matter, and how they are embedded in a more fundamental theory of nature. However, most searches for HNLs that aim for discovery do not rely on such subtle effects.
Hence, while more accurate modelling is clearly needed for precision studies after a discovery, 
the combined sets of benchmarks \eqref{OldBenchmarks} and \eqref{NewBenchmarks}  can describe a wide range of signatures commonly used in searches for HNLs that interact with SM gauge bosons through their mixing with ordinary neutrinos. It would be highly desirable to define a similar set of benchmarks for models with an extended field content, but this remains challenging due to the wealth of possible extensions of the SM and their vastly different phenomenologies. 

\paragraph{Acknowledgments.} This work was prepared for the HNL Working Group of CERN's Feebly Interacting Particles Physics Centre under the umbrella of the PBC initiative. We thank all members of the working group, namely 
Alexey Boyarsky,
Pilar Hernandez, 
Gaia Lanfranchi, 
Silvia Pascoli, and
Mikhail Shaposhnikov, 
for the discussions that led to the definition of the two new benchmarks.
We thank Jean-Loup Tastet and Albert de Roeck for feedback on the manuscript.
MaD and JK would also like to thank Stefan Antusch and Jan Hajer for helpful discussions on HNL oscillations.
J.K. acknowledges the support of the Fonds de la Recherche Scientifique - FNRS under Grant No. 4.4512.10. JLP acknowledges support from Generalitat Valenciana through the plan GenT program (CIDEGENT/2018/019), the European Union Horizon 2020 research and innovation programme under the Marie Sklodowska-Curie grant agreement No 860881-HIDDeN, and the Spanish Ministerio de Ciencia e Innovacion project PID2020-113644GB-I00.

\bibliographystyle{JHEP}
\bibliography{references.bib}{}

\providecommand{\href}[2]{#2}\begingroup\raggedright\begin{thebibliography}{10}

\bibitem{Beacham:2019nyx}
J.~Beacham et~al., \emph{{Physics Beyond Colliders at CERN: Beyond the Standard
  Model Working Group Report}},
  \href{https://doi.org/10.1088/1361-6471/ab4cd2}{\emph{J. Phys. G} {\bfseries
  47} (2020) 010501} [\href{https://arxiv.org/abs/1901.09966}{{\ttfamily
  1901.09966}}].

\bibitem{Agrawal:2021dbo}
P.~Agrawal et~al., \emph{{Feebly-interacting particles: FIPs 2020 workshop
  report}}, \href{https://doi.org/10.1140/epjc/s10052-021-09703-7}{\emph{Eur.
  Phys. J. C} {\bfseries 81} (2021) 1015}
  [\href{https://arxiv.org/abs/2102.12143}{{\ttfamily 2102.12143}}].

\bibitem{FCC:2018evy}
{\scshape FCC} collaboration, A.~Abada et~al., \emph{{FCC-ee: The Lepton
  Collider}: {Future Circular Collider Conceptual Design Report Volume 2}},
  \href{https://doi.org/10.1140/epjst/e2019-900045-4}{\emph{Eur. Phys. J. ST}
  {\bfseries 228} (2019) 261}.

\bibitem{CEPCStudyGroup:2018ghi}
{\scshape CEPC Study Group} collaboration, M.~Dong et~al., \emph{{CEPC
  Conceptual Design Report: Volume 2 - Physics \& Detector}},
  \href{https://arxiv.org/abs/1811.10545}{{\ttfamily 1811.10545}}.

\bibitem{Abdullahi:2022jlv}
A.~M. Abdullahi et~al., \emph{{The Present and Future Status of Heavy Neutral
  Leptons}},  in \emph{{2022 Snowmass Summer Study}}, 3, 2022,
  \href{https://arxiv.org/abs/2203.08039}{{\ttfamily 2203.08039}}.

\bibitem{Minkowski:1977sc}
P.~Minkowski, \emph{{$\mu \to e\gamma$ at a Rate of One Out of $10^{9}$ Muon
  Decays?}}, \href{https://doi.org/10.1016/0370-2693(77)90435-X}{\emph{Phys.
  Lett. B} {\bfseries 67} (1977) 421}.

\bibitem{Glashow:1979nm}
S.~L. Glashow, \emph{{The Future of Elementary Particle Physics}},
  \href{https://doi.org/10.1007/978-1-4684-7197-7_15}{\emph{NATO Sci. Ser. B}
  {\bfseries 61} (1980) 687}.

\bibitem{Gell-Mann:1979vob}
M.~Gell-Mann, P.~Ramond and R.~Slansky, \emph{{Complex Spinors and Unified
  Theories}}, {\emph{Conf. Proc. C} {\bfseries 790927} (1979) 315}
  [\href{https://arxiv.org/abs/1306.4669}{{\ttfamily 1306.4669}}].

\bibitem{Mohapatra:1979ia}
R.~N. Mohapatra and G.~Senjanovic, \emph{{Neutrino Mass and Spontaneous Parity
  Nonconservation}},
  \href{https://doi.org/10.1103/PhysRevLett.44.912}{\emph{Phys. Rev. Lett.}
  {\bfseries 44} (1980) 912}.

\bibitem{Yanagida:1980xy}
T.~Yanagida, \emph{{Horizontal Symmetry and Masses of Neutrinos}},
  \href{https://doi.org/10.1143/PTP.64.1103}{\emph{Prog. Theor. Phys.}
  {\bfseries 64} (1980) 1103}.

\bibitem{Schechter:1980gr}
J.~Schechter and J.~W.~F. Valle, \emph{{Neutrino Masses in SU(2) x U(1)
  Theories}}, \href{https://doi.org/10.1103/PhysRevD.22.2227}{\emph{Phys. Rev.
  D} {\bfseries 22} (1980) 2227}.

\bibitem{Canetti:2012zc}
L.~Canetti, M.~Drewes and M.~Shaposhnikov, \emph{{Matter and Antimatter in the
  Universe}}, \href{https://doi.org/10.1088/1367-2630/14/9/095012}{\emph{New J.
  Phys.} {\bfseries 14} (2012) 095012}
  [\href{https://arxiv.org/abs/1204.4186}{{\ttfamily 1204.4186}}].

\bibitem{Fukugita:1986hr}
M.~Fukugita and T.~Yanagida, \emph{{Baryogenesis Without Grand Unification}},
  \href{https://doi.org/10.1016/0370-2693(86)91126-3}{\emph{Phys. Lett. B}
  {\bfseries 174} (1986) 45}.

\bibitem{Dodelson:1993je}
S.~Dodelson and L.~M. Widrow, \emph{{Sterile-neutrinos as dark matter}},
  \href{https://doi.org/10.1103/PhysRevLett.72.17}{\emph{Phys. Rev. Lett.}
  {\bfseries 72} (1994) 17}
  [\href{https://arxiv.org/abs/hep-ph/9303287}{{\ttfamily hep-ph/9303287}}].

\bibitem{Deppisch:2015qwa}
F.~F. Deppisch, P.~S. Bhupal~Dev and A.~Pilaftsis, \emph{{Neutrinos and
  Collider Physics}},
  \href{https://doi.org/10.1088/1367-2630/17/7/075019}{\emph{New J. Phys.}
  {\bfseries 17} (2015) 075019}
  [\href{https://arxiv.org/abs/1502.06541}{{\ttfamily 1502.06541}}].

\bibitem{Anamiati:2016uxp}
G.~Anamiati, M.~Hirsch and E.~Nardi, \emph{{Quasi-Dirac neutrinos at the LHC}},
  \href{https://doi.org/10.1007/JHEP10(2016)010}{\emph{JHEP} {\bfseries 10}
  (2016) 010} [\href{https://arxiv.org/abs/1607.05641}{{\ttfamily
  1607.05641}}].

\bibitem{SHiP:2018xqw}
{\scshape SHiP} collaboration, C.~Ahdida et~al., \emph{{Sensitivity of the SHiP
  experiment to Heavy Neutral Leptons}},
  \href{https://doi.org/10.1007/JHEP04(2019)077}{\emph{JHEP} {\bfseries 04}
  (2019) 077} [\href{https://arxiv.org/abs/1811.00930}{{\ttfamily
  1811.00930}}].

\bibitem{Drewes:2018gkc}
M.~Drewes, J.~Hajer, J.~Klaric and G.~Lanfranchi, \emph{{NA62 sensitivity to
  heavy neutral leptons in the low scale seesaw model}},
  \href{https://doi.org/10.1007/JHEP07(2018)105}{\emph{JHEP} {\bfseries 07}
  (2018) 105} [\href{https://arxiv.org/abs/1801.04207}{{\ttfamily
  1801.04207}}].

\bibitem{Tastet:2021ygq}
J.-L. Tastet, O.~Ruchayskiy and I.~Timiryasov, \emph{{Why interpretation
  matters for BSM searches: a case study with Heavy Neutral Leptons at ATLAS}},
  \href{https://doi.org/10.22323/1.398.0703}{\emph{PoS} {\bfseries EPS-HEP2021}
  (2022) 703} [\href{https://arxiv.org/abs/2110.11907}{{\ttfamily
  2110.11907}}].

\bibitem{Atre:2009rg}
A.~Atre, T.~Han, S.~Pascoli and B.~Zhang, \emph{{The Search for Heavy Majorana
  Neutrinos}}, \href{https://doi.org/10.1088/1126-6708/2009/05/030}{\emph{JHEP}
  {\bfseries 05} (2009) 030} [\href{https://arxiv.org/abs/0901.3589}{{\ttfamily
  0901.3589}}].

\bibitem{Drewes:2013gca}
M.~Drewes, \emph{{The Phenomenology of Right Handed Neutrinos}},
  \href{https://doi.org/10.1142/S0218301313300191}{\emph{Int. J. Mod. Phys.}
  {\bfseries E22} (2013) 1330019}
  [\href{https://arxiv.org/abs/1303.6912}{{\ttfamily 1303.6912}}].

\bibitem{Antusch:2016ejd}
S.~Antusch, E.~Cazzato and O.~Fischer, \emph{{Sterile neutrino searches at
  future $e^-e^+$, $pp$, and $e^-p$ colliders}},
  \href{https://doi.org/10.1142/S0217751X17500786}{\emph{Int. J. Mod. Phys. A}
  {\bfseries 32} (2017) 1750078}
  [\href{https://arxiv.org/abs/1612.02728}{{\ttfamily 1612.02728}}].

\bibitem{Chun:2017spz}
E.~J. Chun et~al., \emph{{Probing Leptogenesis}},
  \href{https://doi.org/10.1142/S0217751X18420058}{\emph{Int. J. Mod. Phys. A}
  {\bfseries 33} (2018) 1842005}
  [\href{https://arxiv.org/abs/1711.02865}{{\ttfamily 1711.02865}}].

\bibitem{Cai:2017mow}
Y.~Cai, T.~Han, T.~Li and R.~Ruiz, \emph{{Lepton Number Violation: Seesaw
  Models and Their Collider Tests}},
  \href{https://doi.org/10.3389/fphy.2018.00040}{\emph{Front.in Phys.}
  {\bfseries 6} (2018) 40} [\href{https://arxiv.org/abs/1711.02180}{{\ttfamily
  1711.02180}}].

\bibitem{Tastet:2021vwp}
J.-L. Tastet, O.~Ruchayskiy and I.~Timiryasov, \emph{{Reinterpreting the ATLAS
  bounds on heavy neutral leptons in a realistic neutrino oscillation model}},
  \href{https://doi.org/10.1007/JHEP12(2021)182}{\emph{JHEP} {\bfseries 12}
  (2021) 182} [\href{https://arxiv.org/abs/2107.12980}{{\ttfamily
  2107.12980}}].

\bibitem{ATLAS:2022atq}
{\scshape ATLAS} collaboration, \emph{{Search for heavy neutral leptons in
  decays of $W$ bosons using a dilepton displaced vertex in $\sqrt{s}=13$ TeV
  $pp$ collisions with the ATLAS detector}},
  \href{https://arxiv.org/abs/2204.11988}{{\ttfamily 2204.11988}}.

\bibitem{Esteban:2020cvm}
I.~Esteban, M.~C. Gonzalez-Garcia, M.~Maltoni, T.~Schwetz and A.~Zhou,
  \emph{{The fate of hints: updated global analysis of three-flavor neutrino
  oscillations}}, \href{https://doi.org/10.1007/JHEP09(2020)178}{\emph{JHEP}
  {\bfseries 09} (2020) 178}
  [\href{https://arxiv.org/abs/2007.14792}{{\ttfamily 2007.14792}}].

\bibitem{King:2017guk}
S.~F. King, \emph{{Unified Models of Neutrinos, Flavour and CP Violation}},
  \href{https://doi.org/10.1016/j.ppnp.2017.01.003}{\emph{Prog. Part. Nucl.
  Phys.} {\bfseries 94} (2017) 217}
  [\href{https://arxiv.org/abs/1701.04413}{{\ttfamily 1701.04413}}].

\bibitem{Xing:2020ijf}
Z.-z. Xing, \emph{{Flavor structures of charged fermions and massive
  neutrinos}}, \href{https://doi.org/10.1016/j.physrep.2020.02.001}{\emph{Phys.
  Rept.} {\bfseries 854} (2020) 1}
  [\href{https://arxiv.org/abs/1909.09610}{{\ttfamily 1909.09610}}].

\bibitem{Akhmedov:1998qx}
E.~K. Akhmedov, V.~A. Rubakov and A.~{\relax Yu}. Smirnov, \emph{{Baryogenesis
  via neutrino oscillations}},
  \href{https://doi.org/10.1103/PhysRevLett.81.1359}{\emph{Phys. Rev. Lett.}
  {\bfseries 81} (1998) 1359}
  [\href{https://arxiv.org/abs/hep-ph/9803255}{{\ttfamily hep-ph/9803255}}].

\bibitem{Pilaftsis:2003gt}
A.~Pilaftsis and T.~E.~J. Underwood, \emph{{Resonant leptogenesis}},
  \href{https://doi.org/10.1016/j.nuclphysb.2004.05.029}{\emph{Nucl. Phys. B}
  {\bfseries 692} (2004) 303}
  [\href{https://arxiv.org/abs/hep-ph/0309342}{{\ttfamily hep-ph/0309342}}].

\bibitem{Asaka:2005pn}
T.~Asaka and M.~Shaposhnikov, \emph{{The $\nu$MSM, dark matter and baryon
  asymmetry of the universe}},
  \href{https://doi.org/10.1016/j.physletb.2005.06.020}{\emph{Phys. Lett. B}
  {\bfseries 620} (2005) 17}
  [\href{https://arxiv.org/abs/hep-ph/0505013}{{\ttfamily hep-ph/0505013}}].

\bibitem{Davidson:2006tg}
S.~Davidson, G.~Isidori and A.~Strumia, \emph{{The Smallest neutrino mass}},
  \href{https://doi.org/10.1016/j.physletb.2007.01.015}{\emph{Phys. Lett. B}
  {\bfseries 646} (2007) 100}
  [\href{https://arxiv.org/abs/hep-ph/0611389}{{\ttfamily hep-ph/0611389}}].

\bibitem{Shaposhnikov:2006nn}
M.~Shaposhnikov, \emph{{A Possible symmetry of the nuMSM}},
  \href{https://doi.org/10.1016/j.nuclphysb.2006.11.003}{\emph{Nucl. Phys. B}
  {\bfseries 763} (2007) 49}
  [\href{https://arxiv.org/abs/hep-ph/0605047}{{\ttfamily hep-ph/0605047}}].

\bibitem{Kersten:2007vk}
J.~Kersten and A.~Y. Smirnov, \emph{{Right-Handed Neutrinos at CERN LHC and the
  Mechanism of Neutrino Mass Generation}},
  \href{https://doi.org/10.1103/PhysRevD.76.073005}{\emph{Phys. Rev. D}
  {\bfseries 76} (2007) 073005}
  [\href{https://arxiv.org/abs/0705.3221}{{\ttfamily 0705.3221}}].

\bibitem{Moffat:2017feq}
K.~Moffat, S.~Pascoli and C.~Weiland, \emph{{Equivalence between massless
  neutrinos and lepton number conservation in fermionic singlet extensions of
  the Standard Model}},  \href{https://arxiv.org/abs/1712.07611}{{\ttfamily
  1712.07611}}.

\bibitem{Asaka:2005an}
T.~Asaka, S.~Blanchet and M.~Shaposhnikov, \emph{{The nuMSM, dark matter and
  neutrino masses}},
  \href{https://doi.org/10.1016/j.physletb.2005.09.070}{\emph{Phys. Lett. B}
  {\bfseries 631} (2005) 151}
  [\href{https://arxiv.org/abs/hep-ph/0503065}{{\ttfamily hep-ph/0503065}}].

\bibitem{Wyler:1982dd}
D.~Wyler and L.~Wolfenstein, \emph{{Massless Neutrinos in Left-Right Symmetric
  Models}}, \href{https://doi.org/10.1016/0550-3213(83)90482-0}{\emph{Nucl.
  Phys.} {\bfseries B218} (1983) 205}.

\bibitem{Mohapatra:1986aw}
R.~N. Mohapatra, \emph{{Mechanism for Understanding Small Neutrino Mass in
  Superstring Theories}},
  \href{https://doi.org/10.1103/PhysRevLett.56.561}{\emph{Phys. Rev. Lett.}
  {\bfseries 56} (1986) 561}.

\bibitem{Mohapatra:1986bd}
R.~N. Mohapatra and J.~W.~F. Valle, \emph{{Neutrino Mass and Baryon Number
  Nonconservation in Superstring Models}},
  \href{https://doi.org/10.1103/PhysRevD.34.1642}{\emph{Phys. Rev. D}
  {\bfseries 34} (1986) 1642}.

\bibitem{Bernabeu:1987gr}
J.~Bernabeu, A.~Santamaria, J.~Vidal, A.~Mendez and J.~W.~F. Valle,
  \emph{{Lepton Flavor Nonconservation at High-Energies in a Superstring
  Inspired Standard Model}},
  \href{https://doi.org/10.1016/0370-2693(87)91100-2}{\emph{Phys. Lett. B}
  {\bfseries 187} (1987) 303}.

\bibitem{Branco:1988ex}
G.~C. Branco, W.~Grimus and L.~Lavoura, \emph{{The Seesaw Mechanism in the
  Presence of a Conserved Lepton Number}},
  \href{https://doi.org/10.1016/0550-3213(89)90304-0}{\emph{Nucl. Phys. B}
  {\bfseries 312} (1989) 492}.

\bibitem{Akhmedov:1995ip}
E.~K. Akhmedov, M.~Lindner, E.~Schnapka and J.~W.~F. Valle, \emph{{Left-right
  symmetry breaking in NJL approach}},
  \href{https://doi.org/10.1016/0370-2693(95)01504-3}{\emph{Phys. Lett.}
  {\bfseries B368} (1996) 270}
  [\href{https://arxiv.org/abs/hep-ph/9507275}{{\ttfamily hep-ph/9507275}}].

\bibitem{Akhmedov:1995vm}
E.~K. Akhmedov, M.~Lindner, E.~Schnapka and J.~W.~F. Valle, \emph{{Dynamical
  left-right symmetry breaking}},
  \href{https://doi.org/10.1103/PhysRevD.53.2752}{\emph{Phys. Rev.} {\bfseries
  D53} (1996) 2752} [\href{https://arxiv.org/abs/hep-ph/9509255}{{\ttfamily
  hep-ph/9509255}}].

\bibitem{Gavela:2009cd}
M.~B. Gavela, T.~Hambye, D.~Hernandez and P.~Hernandez, \emph{{Minimal Flavour
  Seesaw Models}},
  \href{https://doi.org/10.1088/1126-6708/2009/09/038}{\emph{JHEP} {\bfseries
  09} (2009) 038} [\href{https://arxiv.org/abs/0906.1461}{{\ttfamily
  0906.1461}}].

\bibitem{Hernandez:2016kel}
P.~Hern\'andez, M.~Kekic, J.~L\'opez-Pav\'on, J.~Racker and J.~Salvado,
  \emph{{Testable Baryogenesis in Seesaw Models}},
  \href{https://doi.org/10.1007/JHEP08(2016)157}{\emph{JHEP} {\bfseries 08}
  (2016) 157} [\href{https://arxiv.org/abs/1606.06719}{{\ttfamily
  1606.06719}}].

\bibitem{Drewes:2016jae}
M.~Drewes, B.~Garbrecht, D.~Gueter and J.~Klaric, \emph{{Testing the low scale
  seesaw and leptogenesis}},
  \href{https://doi.org/10.1007/JHEP08(2017)018}{\emph{JHEP} {\bfseries 08}
  (2017) 018} [\href{https://arxiv.org/abs/1609.09069}{{\ttfamily
  1609.09069}}].

\bibitem{Bezrukov:2012sa}
F.~Bezrukov, M.~Y. Kalmykov, B.~A. Kniehl and M.~Shaposhnikov, \emph{{Higgs
  Boson Mass and New Physics}},
  \href{https://doi.org/10.1007/JHEP10(2012)140}{\emph{JHEP} {\bfseries 10}
  (2012) 140} [\href{https://arxiv.org/abs/1205.2893}{{\ttfamily 1205.2893}}].

\bibitem{Klaric:2020phc}
J.~Klari\'c, M.~Shaposhnikov and I.~Timiryasov, \emph{{Uniting Low-Scale
  Leptogenesis Mechanisms}},
  \href{https://doi.org/10.1103/PhysRevLett.127.111802}{\emph{Phys. Rev. Lett.}
  {\bfseries 127} (2021) 111802}
  [\href{https://arxiv.org/abs/2008.13771}{{\ttfamily 2008.13771}}].

\bibitem{Klaric:2021cpi}
J.~Klari\'c, M.~Shaposhnikov and I.~Timiryasov, \emph{{Reconciling resonant
  leptogenesis and baryogenesis via neutrino oscillations}},
  \href{https://doi.org/10.1103/PhysRevD.104.055010}{\emph{Phys. Rev. D}
  {\bfseries 104} (2021) 055010}
  [\href{https://arxiv.org/abs/2103.16545}{{\ttfamily 2103.16545}}].

\bibitem{Hernandez:2022ivz}
P.~Hernandez, J.~Lopez-Pavon, N.~Rius and S.~Sandner, \emph{{Bounds on
  right-handed neutrino parameters from observable leptogenesis}},
  \href{https://arxiv.org/abs/2207.01651}{{\ttfamily 2207.01651}}.

\bibitem{Drewes:2019byd}
M.~Drewes, J.~Klari\'c and P.~Klose, \emph{{On lepton number violation in heavy
  neutrino decays at colliders}},
  \href{https://doi.org/10.1007/JHEP11(2019)032}{\emph{JHEP} {\bfseries 11}
  (2019) 032} [\href{https://arxiv.org/abs/1907.13034}{{\ttfamily
  1907.13034}}].

\bibitem{Das:2017hmg}
A.~Das, P.~S.~B. Dev and R.~N. Mohapatra, \emph{{Same Sign versus Opposite Sign
  Dileptons as a Probe of Low Scale Seesaw Mechanisms}},
  \href{https://doi.org/10.1103/PhysRevD.97.015018}{\emph{Phys. Rev. D}
  {\bfseries 97} (2018) 015018}
  [\href{https://arxiv.org/abs/1709.06553}{{\ttfamily 1709.06553}}].

\bibitem{Abada:2019bac}
A.~Abada, C.~Hati, X.~Marcano and A.~M. Teixeira, \emph{{Interference effects
  in LNV and LFV semileptonic decays: the Majorana hypothesis}},
  \href{https://doi.org/10.1007/JHEP09(2019)017}{\emph{JHEP} {\bfseries 09}
  (2019) 017} [\href{https://arxiv.org/abs/1904.05367}{{\ttfamily
  1904.05367}}].

\bibitem{Cvetic:2015ura}
G.~Cvetic, C.~S. Kim, R.~Kogerler and J.~Zamora-Saa, \emph{{Oscillation of
  heavy sterile neutrino in decay of $B \to \mu e \pi$}},
  \href{https://doi.org/10.1103/PhysRevD.92.013015}{\emph{Phys. Rev. D}
  {\bfseries 92} (2015) 013015}
  [\href{https://arxiv.org/abs/1505.04749}{{\ttfamily 1505.04749}}].

\bibitem{Antusch:2017ebe}
S.~Antusch, E.~Cazzato and O.~Fischer, \emph{{Resolvable heavy
  neutrino\textendash{}antineutrino oscillations at colliders}},
  \href{https://doi.org/10.1142/S0217732319500615}{\emph{Mod. Phys. Lett. A}
  {\bfseries 34} (2019) 1950061}
  [\href{https://arxiv.org/abs/1709.03797}{{\ttfamily 1709.03797}}].

\bibitem{Cvetic:2018elt}
G.~Cveti\v{c}, A.~Das and J.~Zamora-Sa\'a, \emph{{Probing heavy neutrino
  oscillations in rare $W$ boson decays}},
  \href{https://doi.org/10.1088/1361-6471/ab1212}{\emph{J. Phys. G} {\bfseries
  46} (2019) 075002} [\href{https://arxiv.org/abs/1805.00070}{{\ttfamily
  1805.00070}}].

\bibitem{Dib:2017iva}
C.~O. Dib, C.~S. Kim and K.~Wang, \emph{{Signatures of Dirac and Majorana
  sterile neutrinos in trilepton events at the LHC}},
  \href{https://doi.org/10.1103/PhysRevD.95.115020}{\emph{Phys. Rev. D}
  {\bfseries 95} (2017) 115020}
  [\href{https://arxiv.org/abs/1703.01934}{{\ttfamily 1703.01934}}].

\bibitem{Arbelaez:2017zqq}
C.~Arbela\'ez, C.~Dib, I.~Schmidt and J.~C. Vasquez, \emph{{Probing the Dirac
  or Majorana nature of the Heavy Neutrinos in pure leptonic decays at the
  LHC}}, \href{https://doi.org/10.1103/PhysRevD.97.055011}{\emph{Phys. Rev. D}
  {\bfseries 97} (2018) 055011}
  [\href{https://arxiv.org/abs/1712.08704}{{\ttfamily 1712.08704}}].

\bibitem{Balantekin:2018ukw}
A.~B. Balantekin, A.~de~Gouv\^ea and B.~Kayser, \emph{{Addressing the Majorana
  vs. Dirac Question with Neutrino Decays}},
  \href{https://doi.org/10.1016/j.physletb.2018.11.068}{\emph{Phys. Lett. B}
  {\bfseries 789} (2019) 488}
  [\href{https://arxiv.org/abs/1808.10518}{{\ttfamily 1808.10518}}].

\bibitem{Hernandez:2018cgc}
P.~Hern\'andez, J.~Jones-P\'erez and O.~Suarez-Navarro, \emph{{Majorana vs
  Pseudo-Dirac Neutrinos at the ILC}},
  \href{https://doi.org/10.1140/epjc/s10052-019-6728-1}{\emph{Eur. Phys. J. C}
  {\bfseries 79} (2019) 220}
  [\href{https://arxiv.org/abs/1810.07210}{{\ttfamily 1810.07210}}].

\bibitem{Tastet:2019nqj}
J.-L. Tastet and I.~Timiryasov, \emph{{Dirac vs. Majorana HNLs (and their
  oscillations) at SHiP}},
  \href{https://doi.org/10.1007/JHEP04(2020)005}{\emph{JHEP} {\bfseries 04}
  (2020) 005} [\href{https://arxiv.org/abs/1912.05520}{{\ttfamily
  1912.05520}}].

\bibitem{Blondel:2021mss}
A.~Blondel, A.~de~Gouv\^ea and B.~Kayser, \emph{{Z-boson decays into Majorana
  or Dirac heavy neutrinos}},
  \href{https://doi.org/10.1103/PhysRevD.104.055027}{\emph{Phys. Rev. D}
  {\bfseries 104} (2021) 055027}
  [\href{https://arxiv.org/abs/2105.06576}{{\ttfamily 2105.06576}}].

\bibitem{Dib:2016wge}
C.~O. Dib, C.~S. Kim, K.~Wang and J.~Zhang, \emph{{Distinguishing
  Dirac/Majorana Sterile Neutrinos at the LHC}},
  \href{https://doi.org/10.1103/PhysRevD.94.013005}{\emph{Phys. Rev. D}
  {\bfseries 94} (2016) 013005}
  [\href{https://arxiv.org/abs/1605.01123}{{\ttfamily 1605.01123}}].

\bibitem{Alimena:2022hfr}
J.~Alimena et~al., \emph{{Searches for Long-Lived Particles at the Future
  FCC-ee}},  \href{https://arxiv.org/abs/2203.05502}{{\ttfamily 2203.05502}}.

\bibitem{Antusch:2017pkq}
S.~Antusch, E.~Cazzato, M.~Drewes, O.~Fischer, B.~Garbrecht, D.~Gueter et~al.,
  \emph{{Probing Leptogenesis at Future Colliders}},
  \href{https://doi.org/10.1007/JHEP09(2018)124}{\emph{JHEP} {\bfseries 09}
  (2018) 124} [\href{https://arxiv.org/abs/1710.03744}{{\ttfamily
  1710.03744}}].

\bibitem{Boyanovsky:2014una}
D.~Boyanovsky, \emph{{Nearly degenerate heavy sterile neutrinos in cascade
  decay: mixing and oscillations}},
  \href{https://doi.org/10.1103/PhysRevD.90.105024}{\emph{Phys. Rev. D}
  {\bfseries 90} (2014) 105024}
  [\href{https://arxiv.org/abs/1409.4265}{{\ttfamily 1409.4265}}].

\bibitem{Antusch:2020pnn}
S.~Antusch and J.~Rosskopp, \emph{{Heavy Neutrino-Antineutrino Oscillations in
  Quantum Field Theory}},
  \href{https://doi.org/10.1007/JHEP03(2021)170}{\emph{JHEP} {\bfseries 03}
  (2021) 170} [\href{https://arxiv.org/abs/2012.05763}{{\ttfamily
  2012.05763}}].

\bibitem{Alva:2014gxa}
D.~Alva, T.~Han and R.~Ruiz, \emph{{Heavy Majorana neutrinos from $W\gamma$
  fusion at hadron colliders}},
  \href{https://doi.org/10.1007/JHEP02(2015)072}{\emph{JHEP} {\bfseries 02}
  (2015) 072} [\href{https://arxiv.org/abs/1411.7305}{{\ttfamily 1411.7305}}].

\bibitem{Degrande:2016aje}
C.~Degrande, O.~Mattelaer, R.~Ruiz and J.~Turner, \emph{{Fully-Automated
  Precision Predictions for Heavy Neutrino Production Mechanisms at Hadron
  Colliders}}, \href{https://doi.org/10.1103/PhysRevD.94.053002}{\emph{Phys.
  Rev. D} {\bfseries 94} (2016) 053002}
  [\href{https://arxiv.org/abs/1602.06957}{{\ttfamily 1602.06957}}].

\bibitem{Pascoli:2018heg}
S.~Pascoli, R.~Ruiz and C.~Weiland, \emph{{Heavy neutrinos with dynamic jet
  vetoes: multilepton searches at $ \sqrt{s}=14 $ , 27, and 100 TeV}},
  \href{https://doi.org/10.1007/JHEP06(2019)049}{\emph{JHEP} {\bfseries 06}
  (2019) 049} [\href{https://arxiv.org/abs/1812.08750}{{\ttfamily
  1812.08750}}].

\bibitem{Coloma:2020lgy}
P.~Coloma, E.~Fern\'andez-Mart\'\i{}nez, M.~Gonz\'alez-L\'opez,
  J.~Hern\'andez-Garc\'\i{}a and Z.~Pavlovic, \emph{{GeV-scale neutrinos:
  interactions with mesons and DUNE sensitivity}},
  \href{https://doi.org/10.1140/epjc/s10052-021-08861-y}{\emph{Eur. Phys. J. C}
  {\bfseries 81} (2021) 78} [\href{https://arxiv.org/abs/2007.03701}{{\ttfamily
  2007.03701}}].

\bibitem{Caputo:2017pit}
A.~Caputo, P.~Hernandez, J.~Lopez-Pavon and J.~Salvado, \emph{{The seesaw
  portal in testable models of neutrino masses}},
  \href{https://doi.org/10.1007/JHEP06(2017)112}{\emph{JHEP} {\bfseries 06}
  (2017) 112} [\href{https://arxiv.org/abs/1704.08721}{{\ttfamily
  1704.08721}}].

\bibitem{Casas:2001sr}
J.~A. Casas and A.~Ibarra, \emph{{Oscillating neutrinos and muon ---> e,
  gamma}}, \href{https://doi.org/10.1016/S0550-3213(01)00475-8}{\emph{Nucl.
  Phys.} {\bfseries B618} (2001) 171}
  [\href{https://arxiv.org/abs/hep-ph/0103065}{{\ttfamily hep-ph/0103065}}].

\bibitem{Molinaro:2008rg}
E.~Molinaro and S.~T. Petcov, \emph{{The Interplay Between the 'Low' and 'High'
  Energy CP-Violation in Leptogenesis}},
  \href{https://doi.org/10.1140/epjc/s10052-009-0985-3}{\emph{Eur. Phys. J. C}
  {\bfseries 61} (2009) 93} [\href{https://arxiv.org/abs/0803.4120}{{\ttfamily
  0803.4120}}].

\bibitem{Bondarenko:2021cpc}
K.~Bondarenko, A.~Boyarsky, J.~Klaric, O.~Mikulenko, O.~Ruchayskiy, V.~Syvolap
  et~al., \emph{{An allowed window for heavy neutral leptons below the kaon
  mass}}, \href{https://doi.org/10.1007/JHEP07(2021)193}{\emph{JHEP} {\bfseries
  07} (2021) 193} [\href{https://arxiv.org/abs/2101.09255}{{\ttfamily
  2101.09255}}].

\bibitem{DUNE:2020jqi}
{\scshape DUNE} collaboration, B.~Abi et~al., \emph{{Long-baseline neutrino
  oscillation physics potential of the DUNE experiment}},
  \href{https://doi.org/10.1140/epjc/s10052-020-08456-z}{\emph{Eur. Phys. J. C}
  {\bfseries 80} (2020) 978}
  [\href{https://arxiv.org/abs/2006.16043}{{\ttfamily 2006.16043}}].

\bibitem{DUNE:2020ypp}
{\scshape DUNE} collaboration, B.~Abi et~al., \emph{{Deep Underground Neutrino
  Experiment (DUNE), Far Detector Technical Design Report, Volume II: DUNE
  Physics}},  \href{https://arxiv.org/abs/2002.03005}{{\ttfamily 2002.03005}}.

\bibitem{Caputo:2016ojx}
A.~Caputo, P.~Hernandez, M.~Kekic, J.~L\'opez-Pav\'on and J.~Salvado,
  \emph{{The seesaw path to leptonic CP violation}},
  \href{https://doi.org/10.1140/epjc/s10052-017-4823-8}{\emph{Eur. Phys. J. C}
  {\bfseries 77} (2017) 258}
  [\href{https://arxiv.org/abs/1611.05000}{{\ttfamily 1611.05000}}].

\bibitem{Hyper-Kamiokande:2018ofw}
{\scshape Hyper-Kamiokande} collaboration, K.~Abe et~al.,
  \emph{{Hyper-Kamiokande Design Report}},
  \href{https://arxiv.org/abs/1805.04163}{{\ttfamily 1805.04163}}.

\bibitem{Bezrukov:2005mx}
F.~L. Bezrukov, \emph{{nu MSM-predictions for neutrinoless double beta decay}},
  \href{https://doi.org/10.1103/PhysRevD.72.071303}{\emph{Phys. Rev. D}
  {\bfseries 72} (2005) 071303}
  [\href{https://arxiv.org/abs/hep-ph/0505247}{{\ttfamily hep-ph/0505247}}].

\bibitem{Blennow:2010th}
M.~Blennow, E.~Fernandez-Martinez, J.~Lopez-Pavon and J.~Menendez,
  \emph{{Neutrinoless double beta decay in seesaw models}},
  \href{https://doi.org/10.1007/JHEP07(2010)096}{\emph{JHEP} {\bfseries 07}
  (2010) 096} [\href{https://arxiv.org/abs/1005.3240}{{\ttfamily 1005.3240}}].

\bibitem{Lopez-Pavon:2012yda}
J.~Lopez-Pavon, S.~Pascoli and C.-f. Wong, \emph{{Can heavy neutrinos dominate
  neutrinoless double beta decay?}},
  \href{https://doi.org/10.1103/PhysRevD.87.093007}{\emph{Phys. Rev. D}
  {\bfseries 87} (2013) 093007}
  [\href{https://arxiv.org/abs/1209.5342}{{\ttfamily 1209.5342}}].

\bibitem{Drewes:2016lqo}
M.~Drewes and S.~Eijima, \emph{{Neutrinoless double $\beta$ decay and low scale
  leptogenesis}},
  \href{https://doi.org/10.1016/j.physletb.2016.09.054}{\emph{Phys. Lett. B}
  {\bfseries 763} (2016) 72}
  [\href{https://arxiv.org/abs/1606.06221}{{\ttfamily 1606.06221}}].

\bibitem{Chrzaszcz:2019inj}
M.~Chrzaszcz, M.~Drewes, T.~E. Gonzalo, J.~Harz, S.~Krishnamurthy and
  C.~Weniger, \emph{{A frequentist analysis of three right-handed neutrinos
  with GAMBIT}},  \href{https://arxiv.org/abs/1908.02302}{{\ttfamily
  1908.02302}}.

\bibitem{Abada:2018oly}
A.~Abada, G.~Arcadi, V.~Domcke, M.~Drewes, J.~Klaric and M.~Lucente,
  \emph{{Low-scale leptogenesis with three heavy neutrinos}},
  \href{https://doi.org/10.1007/JHEP01(2019)164}{\emph{JHEP} {\bfseries 01}
  (2019) 164} [\href{https://arxiv.org/abs/1810.12463}{{\ttfamily
  1810.12463}}].

\bibitem{Drewes:2021nqr}
M.~Drewes, Y.~Georis and J.~Klari\'c, \emph{{Mapping the Viable Parameter Space
  for Testable Leptogenesis}},
  \href{https://doi.org/10.1103/PhysRevLett.128.051801}{\emph{Phys. Rev. Lett.}
  {\bfseries 128} (2022) 051801}
  [\href{https://arxiv.org/abs/2106.16226}{{\ttfamily 2106.16226}}].

\end{thebibliography}\endgroup
\end{document}